\documentclass{article}

\usepackage{amsmath}
\usepackage{graphicx}
\usepackage[]{wasysym}
\usepackage{arxiv}
\newcommand{\Var}{\mathrm{Var}}

\usepackage[utf8]{inputenc} % allow utf-8 input
\usepackage[T1]{fontenc}    % use 8-bit T1 fonts
\usepackage{hyperref}       % hyperlinks
\usepackage{url}            % simple URL typesetting
\usepackage{booktabs}       % professional-quality tables
\usepackage{amsfonts}       % blackboard math symbols
\usepackage{nicefrac}       % compact symbols for 1/2, etc.
\usepackage{microtype}      % microtypography
\usepackage{lipsum}

\title{Non-invasive light focusing in scattering media using speckle variance optimization}

\author{
  $^{1}$Antoine Boniface, $^{1,2}$Baptiste Blochet, $^{1,3}$Jonathan Dong $\&$ $^1$Sylvain Gigan \\
 $^1$ Laboratoire Kastler Brossel, Sorbonne Université, École Normale Supérieure–Paris Sciences et Lettres (PSL)\\ Research University, CNRS, Collège de France, 24 rue Lhomond, 75005 Paris, France\\
 $^2$ IBENS, Département de Biologie, Ecole Normale Supérieure, CNRS, Inserm, \\
 PSL Research University, 75005 Paris, France \\
 $^3$ Laboratoire de Physique de l'\'Ecole Normale Sup\'erieure, Universit\'e PSL, CNRS, \\
 Sorbonne Universit\'e, Universit\'e Paris-Diderot, Sorbonne Paris Cit\'e, Paris, France
  }
  
\begin{document}
\maketitle
\begin{abstract}
Optical imaging deep inside scattering media remains a fundamental problem in bio-imaging.  While wavefront shaping has been shown to allow focusing of coherent light at depth,  achieving it non-invasively remains a challenge. Various feedback mechanisms, in particular acoustic or non-linear fluorescence-based, have been put forward for this purpose. Non-invasive focusing at depth on fluorescent objects with linear excitation is, however, still unresolved. Here we report a simple method for focusing inside a scattering medium in an epi-detection geometry with a linear signal: optimizing the spatial variance of low contrast speckle patterns emitted by a set of fluorescent sources. Experimentally, we demonstrate robust and efficient focusing of scattered light on a single source, and show that this variance optimization method is formally equivalent to previous optimization strategies based on two-photon fluorescence. Our technique should generalize to a large variety of incoherent contrast mechanisms and holds interesting prospects for deep bio-imaging.
\end{abstract}

\section*{Introduction}
Disordered media, such as biological tissues, are a major hindrance to retrieving information at depth with light, in particular for imaging. 
Propagation of light is strongly perturbed due to refractive index inhomogeneities. In the multiple scattering regime, ballistic light is exponentially attenuated with depth, and coherent light gives rise to an extended speckle pattern \cite{goodman1976some}. As a consequence, all point scanning or wide field imaging techniques rapidly fails beyond a few hundred microns in tissues. Wavefront shaping techniques have emerged as an extremely effective way to inverse the effect of scattering and focus light to a diffraction-limited spot \cite{rotter2017light}. This is achieved by first measuring a feedback signal from the targeted focal point and then correcting the incident wavefront with a spatial light modulator (SLM). Initially, the feedback was measured with a detector placed behind the scattering medium \cite{vellekoop2007focusing, popoff2010measuring, vellekoop2010exploiting, yaqoob2008optical}, or recovered from a single implanted guide star \cite{Hsieh:10, vellekoop2008demixing}. Using two-photon fluorescence as a feedback, it was shown in \cite{katz2011focusing} that a focused spot could be retrieved, even from an extended object. This was later demonstrated non-invasively and used to scan the focused spot, thanks to the memory effect, in order to image at depth \cite{tang2012superpenetration, katz2014noninvasive}. A wide set of feedback techniques nowadays allows to focus at depth inside complex media \cite{horstmeyer2015guidestar}. Linear fluorescence microscopy remains an inescapable tool in life science \cite{lichtman2005fluorescence, webb2012epi}, allowing superficial layers of a biological sample to be imaged with high resolution and a variety of contrasts. 
However, the general problem of focusing at depth, non-invasively, using linear fluorescence feedback, on extended or multiple targets, remains unsolved. Maximizing the total linear fluorescence results in an extended focus. 

Here, we demonstrate that by choosing as a metric for the wavefront optimization not the total linear fluorescence reflected by the medium but the spatial variance of the fluorescence speckle pattern it reflects, we can successfully generate a single diffraction-limited focus inside the scattering medium. Simply put, our method relies on the fact that a collection of fluorescent sources do generate a speckle, but the speckle contrast is linked to the number of incoherent sources that are added incoherently on the detector. Hence maximizing the variance tends to concentrate the excitation on a single source.  Other methods have very recently been proposed to focus light using linear signal \cite{Stern:19,daniel2019noninvasive}, both exploiting speckle spatial information, however they require the presence of memory effect, thus only works for relatively thin samples. Our method is therefore simpler and more general.
We finally show that once a single bead has been isolated, the centroid of the diffuse spot allows, within a few microns, to localize the target.

\section{Principle}
Contrarily to a common misconception, a fluorescent object, despite being spatially and temporally incoherent, can generate speckle through a complex medium. However, the speckle pattern that one recovers -- that we will refer to as fluorescent speckle by convenience -- is low contrast. First, each fluorescent source, emitting broadband and unpolarized light, generates a speckle whose contrast is decreased: each polarization and spectral band form independent speckles which are summed incoherently \cite{goodman1976some}. Second, speckles from all the fluorescent sources, excited inside the medium, are also added incoherently. The overall contrast of the final speckle is directly linked to the number of independent speckles \cite{goodman1976some}, but only decreases with a mild square root dependence. On the other hand, its intensity scales linearly with the excitation intensity, in the case of linear fluorescence. 

Here, we take advantage of the product of these two by optimizing the linear fluorescence spatial standard deviation $\sigma$ to focus the illumination on a single fluorescent target. We point out here that since spatial variance $\Var$ is just the square of the standard deviation ($\Var=\sigma^2$), we refer for simplicity to variance optimization, but the experimentally-fitted parameter throughout the process is the standard deviation. 

\begin{figure}[htbp]
\centering
\includegraphics[width = 300 pt]{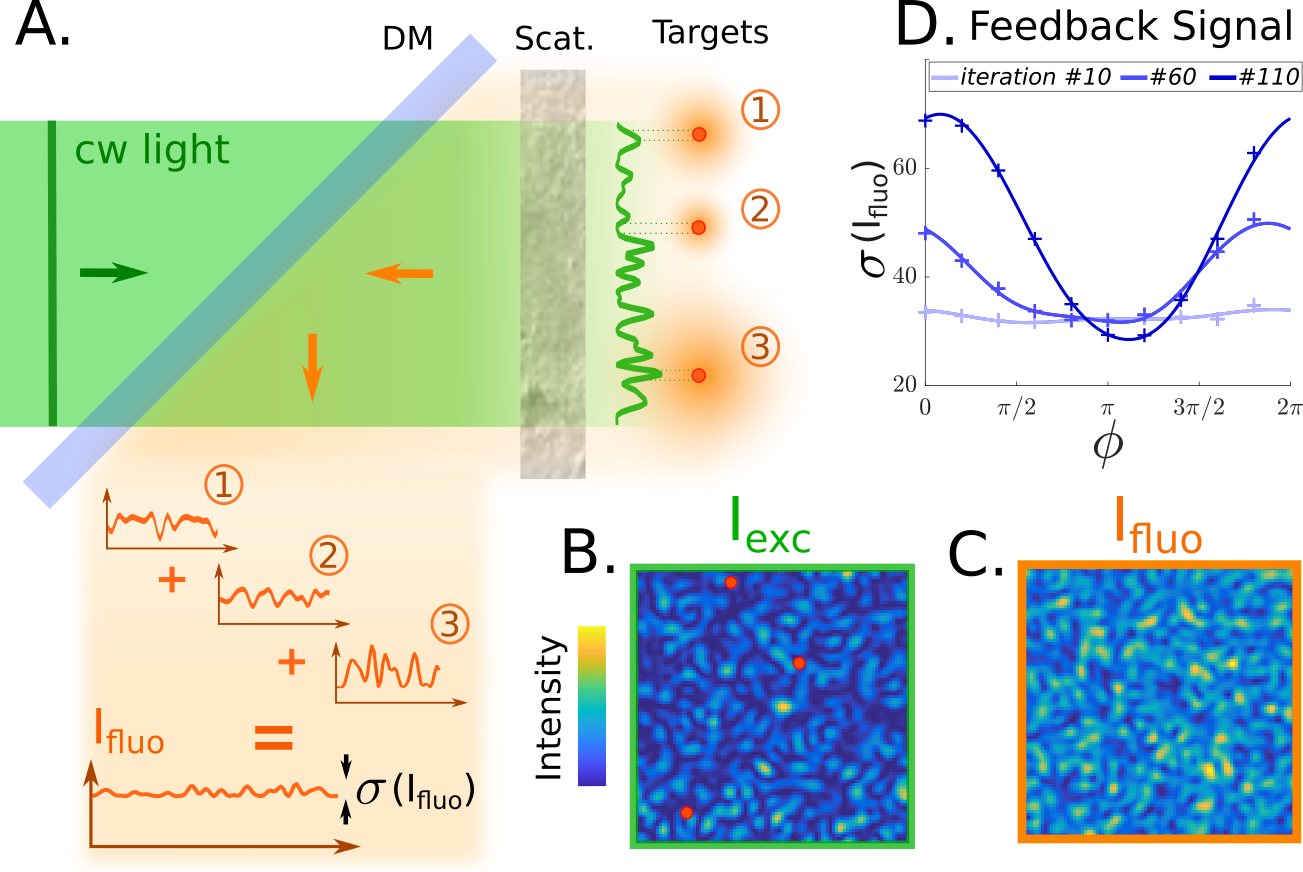}
\caption{Scheme of principle. \textbf{A}. A sparse set of fluorescent sources is excited with a speckle illumination (\textbf{B}, simulation result) and emit linear fluorescence. The resulting epi-detected 2D-signal is a low contrast speckle pattern (\textbf{C}, simulation result). Its spatial standard deviation, $\sigma(I_{fluo})=\sqrt{\Var(I_{fluo})}$, is used as a metric to run the optimization. \textbf{D}. Figures of merit for 3 different iterations. Standard deviation measured data with respect to $\Phi$ are fitted with $\sqrt{(\ref{eq1})}$.}
\label{principle}
\end{figure}

The principle of the method is depicted in Figure\ref{principle}. We excite $N$ fluorescent targets, hidden behind a scattering medium, with a speckle illumination Fig\ref{principle}.B. Linear and isotropic emitted fluorescence forms a low-contrast speckle, $I_{fluo}$ in Fig\ref{principle}.C. The latter is the incoherent sum of the $N$ uncorrelated speckles generated by all the targets, weighted by the excitation intensity each target receives. An important point to emphasize here is that all of these $N$ uncorrelated speckle patterns are intrinsically lowered in contrast due to the broad emission bandwidth.
However, the overall contrast of the fluorescent speckle, $C(I_{fluo})$, is related to the number of excited fluorescent targets and scales as $1/\sqrt{N}$, \cite{goodman1976some}.  
Consequently if one shapes the illumination such that fewer targets are excited, the contrast should increase. In other words, the number of excited targets is closely linked to the contrast of the fluorescent pattern. In particular, $C(I_{fluo})$ is maximum when only a single target is excited while the other $N-1$ receive almost zero intensity.
To enhance the intensity on this target and form a focus with a significant signal-to-background ratio, we also need to maximize the total fluorescence signal. Therefore the metric reads $ C(I_{fluo})\times I_{fluo}$ which corresponds to the standard deviation of the fluorescent speckle: $ \sigma(I_{fluo})$. Using spatial standard deviation as a metric allows us to create a non-linear feedback signal.

Our variance-based wavefront shaping technique consists in optimizing the phase $\Phi$ of each input mode (modulated with the SLM) to maximize $\sigma(I_{fluo})$, and consequently the variance $\Var(I_{fluo})=\sigma(I_{fluo})^2$. 
To have a better understanding of how one mode is optimized, we derive the relation that links the spatial variance, $\Var(I_{fluo}(x,y,\Phi))$, to the excitation intensity $I_{exc}^{(k)}$ on the $k^{th}$ target and the incident phase $\Phi$. We obtain the following equation and give a detailed demonstration in the Supplementary Materials:
\begin{equation}
  \Var(I_{fluo}(x,y,\Phi)) \propto\sum_{k}^{N}I_{exc}^{(k)}(\Phi)^2 \equiv A\sin(2\Phi+\theta_A)+B\sin(\Phi+\theta_B)+C
\label{eq1}
\end{equation}
where $A, \theta_A, B, \theta_B$ and $C$ are constants.

Equation (\ref{eq1}) explicitly highlights that spatial variance introduces a non-linearity of order 2 with respect to the excitation intensity. This justifies that even though the response of the fluorescent targets is a linear signal, its variance generates a non-linear feedback signal. This new optimization scheme enables light focusing on a single fluorescent target, similarly to \cite{katz2014noninvasive} and other works using total two-photon fluorescence as feedback. 
In Figure\ref{principle}.D. we validate our model and show that standard deviation experimental data, $ \sigma(I_{fluo})$, fit well with $\sqrt{eq.(\ref{eq1})}$.

\section{Experimental demonstration}

In the following section we report on the experimental implementation of our variance-based optimization with linear fluorescence. We first describe the optical setup, then detail the optimization algorithm and present experimental results.
\subsection{Experimental setup}
\begin{figure}[htbp]
\centering
\includegraphics[width=300 pt] {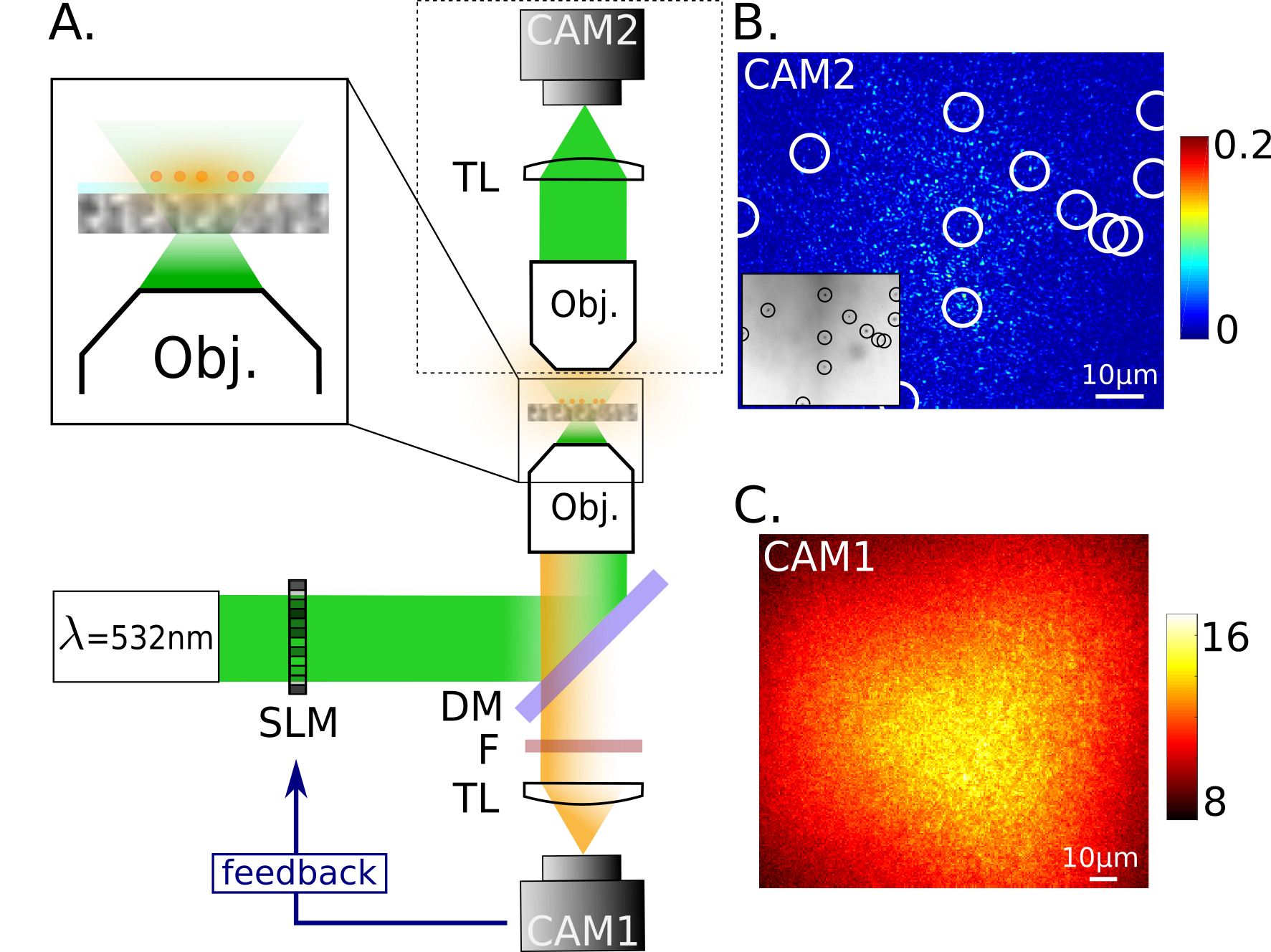}
\caption{Experimental Setup. \textbf{A}. Fluorescence microscopy through few layers of parafilm in epi detection, DM: dichroic mirror, TL: tube lens,  F: filter. The fluorescent speckle (shown in \textbf{C}) is epi-detected on CAM1. Its spatial standard deviation ($std2$(\textbf{C})) is the feedback of our optimization algorithm. A second camera placed in transmission, CAM2, not only monitors the illumination speckle in the plane of the beads (\textbf{B}) but also allows obtaining a bright field image of the beads (\textbf{B} inset). Note that CAM2 is not needed for the optimization but acts as a control camera.}
\label{setup_epi}
\end{figure}

The experimental setup is shown in Fig.\ref{setup_epi}. A cw laser ($\lambda$=532 nm, Coherent Sapphire) is expanded on a phase-only MEMS SLM (Kilo-DM segmented, Boston Micromachines), so that all the 1024 SLM segments are illuminated. The SLM is conjugated to the back focal plane of a microscope objective (Zeiss W "Plan-Apochromat" $20\times$, NA 1.0 ) exciting orange fluorescent beads (Invitrogen™ FluoSpheres, 1.0 $\mu m$) placed behind three layers of parafilm. Complementary measurements allow us to claim that such a scattering medium has virtually no memory effect in the plane of the beads. 
The excitation beam (diameter $<$ 6mm) underfills the illumination objective back aperture (diameter $\simeq$ 20mm) which reduces the actual illumination NA. 
The scattered 1-photon fluorescence emission is collected by the same microscope objective (epi-detection configuration) and detected by a first camera: CAM1 (sCMOS, Hammamatsu ORCA Flash). These recorded fluorescence images are then used to optimize the variance. We use a dichroic mirror (DM) shortpass 550nm (Thorlabs) and filters (F): a 532nm longpass (Semrock) and a 533nm notch (Thorlabs). A second microscope objective (Olympus "MPlan N" $50\times$, NA 0.75) images in transmission the plane of the beads onto a CCD camera (Allied Vision, Manta G-046B) as a passive control. This part of the setup allows us to: (1) get an image of the beads in bright field using white light (MORITEX, MHAB 150W), and (2) monitor the speckle illumination and verify our ability to focus it on a single fluorophore. 

\subsection{Variance optimization algorithm}
The optimization algorithm we use works as follows. For each iteration, we modulate the phase of half of the pixels according to a Hadamard mode (binary basis whose entries are either +1 or -1) on the SLM. One half of the pixels (corresponding to entries "+1", for example) of the selected mode are discretely (with $N_{step}$) modulated in phase from $0$ to $2\pi$. The other half of pixels (corresponding to entries "-1") are not modulated and act as a reference.
For each one of the $N_{step}$, we acquire a fluorescent speckle pattern on CAM1 and calculate its standard deviation. In our experiment we use $N_{step}=8$.
At each iteration, the algorithm finds the phase that maximizes the spatial standard deviation, thus the spatial variance. A new optimum phase mask is then calculated by adding to the previous mask the Hadamard mode with this optimum phase. This mask is applied to the SLM before starting the next iteration. When the full Hadamard basis has been optimized, the algorithm restarts with the first mode of this basis. We can monitor using the control camera how a single focus with high signal-to-background (SBR) is formed. Note that we expect that this method would also be compatible with other optimization strategies proposed to focus in complex media, such as genetic algorithms \cite{Conkey:12}.

\subsection{Focusing light on a single target}
With the optimization algorithm we have detailed in the previous subsection, we report on the results obtained through 5 different optimization procedures. Each time we ran 1500 iterations and used 1024 input Hadamard modes. Our scattering medium is made of 3 layers of parafilm (thickness $\simeq$ 400$\mu m$). The resulting speckle pattern illuminates 12 beads. The latter are dried on a glass coverslip (thickness $\simeq$ 160$\mu m$) and placed on top of the scattering layers of parafilm. These experimental conditions (scattering medium and fluorescent sample) are kept identical for the 5 optimizations.
\begin{figure}[htbp]
\centering
\includegraphics[width=300 pt] {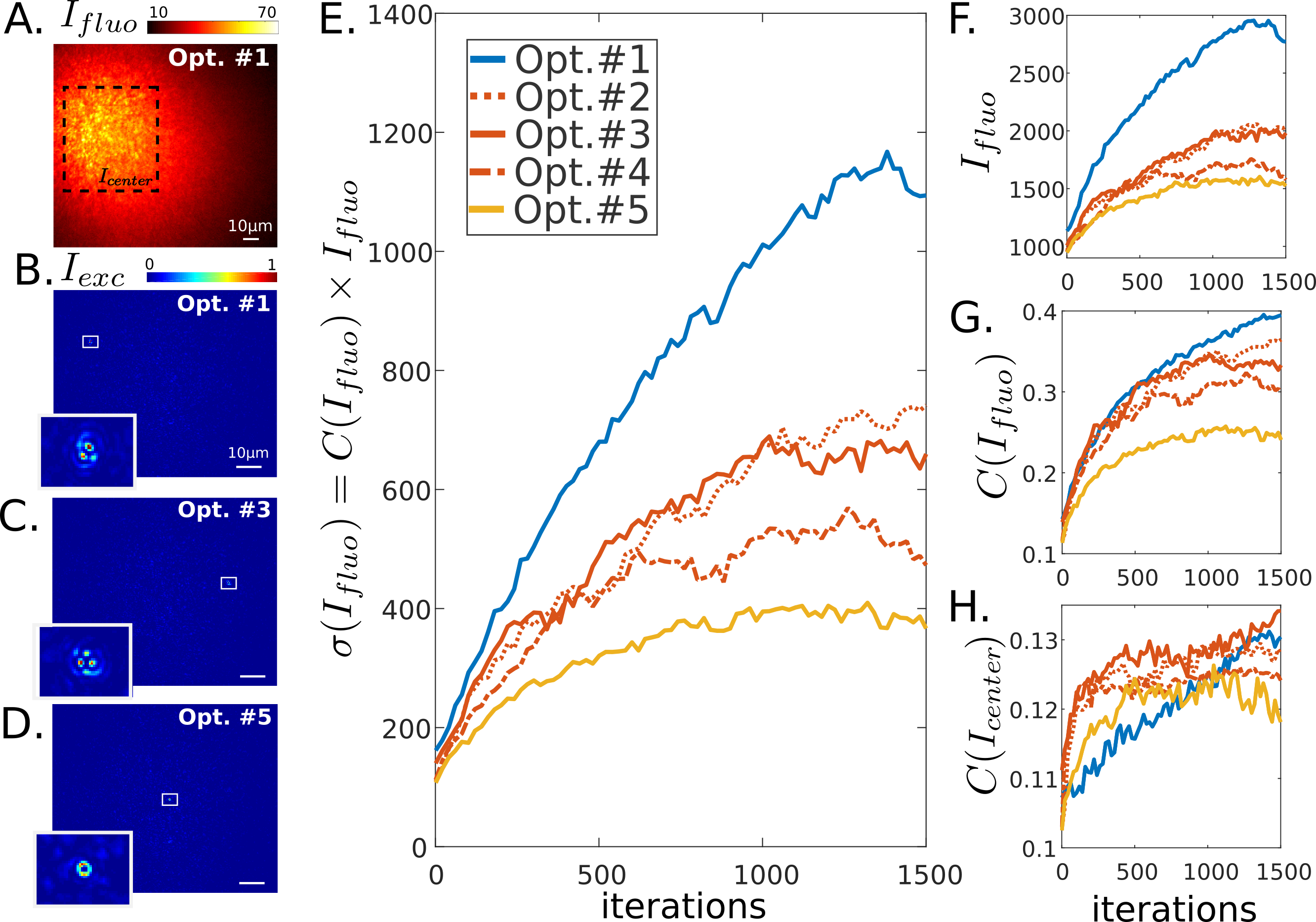}
\caption{Examples of single bead focusing obtained over 5 different optimization procedures. By maximizing the variance of the fluorescent images (graph \textbf{E}), a focus is consistently formed on one of the 12 possible beads. \textbf{A} and \textbf{B} show the fluorescence speckle and the final focus on the bead respectively, for the first optimization (Opt. $\#$1). Two other spatial focii on different beads are also represented (\textbf{C} and \textbf{D}) for different optimization initial conditions. \textbf{F}, \textbf{G} and \textbf{H} represent the evolution of the total intensity, the contrast of the full image and the contrast of the central region of the diffuse spot respectively, and allows to monitor the contributions of each parameter to the overall variance.}
\label{figure_focus_14_03}
\end{figure}

The spatial standard deviation, $\sigma(I_{fluo})$, is estimated throughout all the optimization process from the full image recorded on CAM1. 
Fig.\ref{figure_focus_14_03}.A corresponds to the final fluorescent speckle of Opt. $\#$1 to be compared with Fig.\ref{setup_epi}.C taken before the optimization. It shows that maximizing the variance must be considered at two scales: firstly, on that of speckle grain size, where a contrast enhancement  makes reveals the speckle grains , secondly, on the speckle envelope which tends to shrink when a single bead is selected. Both of these two effects contribute to focus light. 

The corresponding excitation in the plane of the beads (see in Fig.\ref{figure_focus_14_03}.B) is monitored with the control camera placed in transmission (CAM2). 
As it can be seen, the variance enhancement of the fluorescent speckle is achieved by exciting only one target,i.e. focusing the illumination on the bead. 
However, the position of the focus cannot be determined in advance. Additionally, if one changes the initial speckle illumination and/or simply rearrange the order of input modes, another focus may potentially be generated.
Over five realizations, two other spatial focii on different beads are obtained (Fig.\ref{figure_focus_14_03}.C and D).
We observe clearly a single focus spot for each illumination, even though the exact enhancement cannot be precisely determined: focii are indeed distorted because beads themselves absorb and diffract light.

In Fig.\ref{figure_focus_14_03}.E, we plot $\sigma(I_{fluo})$ throughout the whole process. In all cases, it increases continuously and reaches a plateau after $\simeq 1000$ iterations, which is consistent with the number of input modes (1024 SLM pixels). We also note that $C(I_{fluo})$ does not converge to a unique value. The latter is indeed really sensitive to the position of the envelope across the field of view. Furthermore, we estimate, a posteriori, the contrast of a sub-area of the full images recorded on CAM1, denoted $C(I_{center})$, in order to only investigate the contribution of the speckle without its envelope shape. We crop fluorescent images (Fig.\ref{figure_focus_14_03}.A) and set height and width such that it captures all the intensity pixels above 80$\%$ of the maximum intensity of $I_{fluo}$. Since the envelope shrinks throughout the optimization, the cropped area is increasingly small. Contrast of the fluorescent speckle pattern, $C(I_{center})$, significantly increases during the first iterations (Fig.\ref{figure_focus_14_03}.H) and rapidly converges to $C(I_{center})\simeq0.13$ much bellow 1, because collected fluorescence is broadband and unpolarized. At the beginning, the optimization tends to shape the speckle to excite a single bead, which seems to be the best scenario to substantially increase the variance. Once a single focus is obtained, the variance enhancement mainly comes from the total fluorescence enhancement. This leads to a higher SBR of the generated focus (Fig.\ref{figure_focus_14_03}.F). We also report on the performances of our variance-based optimization for a larger number of beads in the Supplementary Materials.

\subsection{Fluorescent targets localization}
At high depth, the memory effect range is too small to form an image by raster scanning the focus over the sample. Nevertheless, the fluorescent speckles after optimization contain spatial information on the position of their emitters.
We take advantage of using a 2D detector (as CAM1), required for the optimization, to extract information on the position of the excited emitter. Indeed when light is focused on a single target, the envelope of the fluorescent speckle is shrunk compared to the initial one, see Fig.\ref{figure_focus_14_03}.A. The observed diffuse spot is simply the propagation of a localized emitted fluorescence through the scattering medium.

\begin{figure}[htbp]
\centering
\includegraphics[width=250 pt] {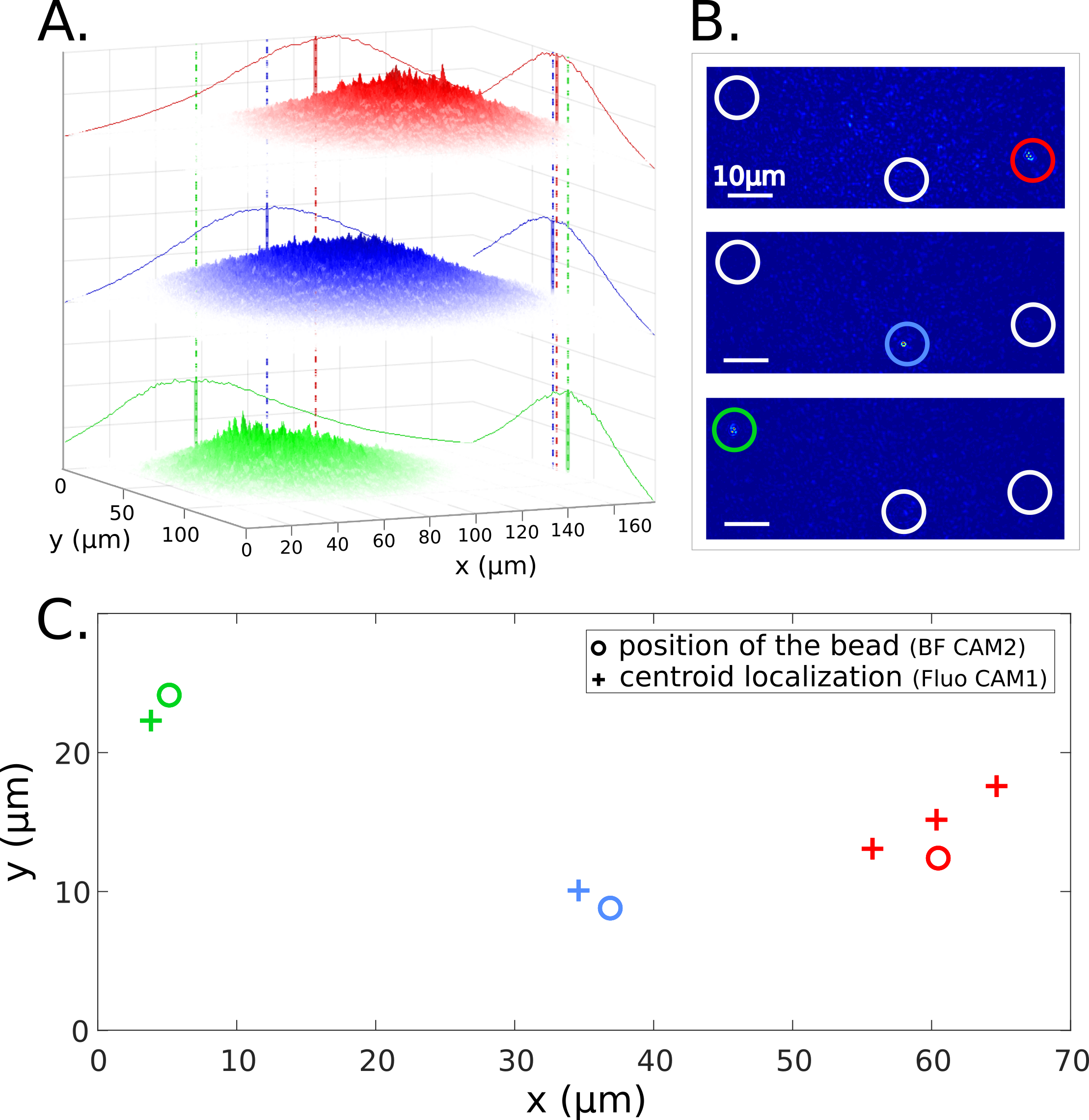}
\caption{Target localization. \textbf{A}. For each of the five realizations shown previously, we determine the centroid of the final epi-detected fluorescent speckle (CAM1). These values are superimposed with the position of the beads on which light is focused (\textbf{B}) retrieved in bright-field illumination in transmission (CAM2). \textbf{C} shows the good agreement, demonstrating the ability to locate the beads qualitatively within a few microns, relatively to each other.}
\label{localizationv2}
\end{figure}

We can estimate the position of the emitter by computing a 2D centroid localization of the fluorescence envelope obtained at the end of the optimization. We actually perform two separate 1D fit, for both directions, on the projected data, as it can be seen in Fig.\ref{localizationv2}.A.
We can notice that the three successive optimizations focusing on the same bead (Opt.$\#2,\#3$ and $\#4$), give rise to similar diffuse spots located in an area whose typical size $\simeq 10 \mu m$. This means that, in our case, this technique would not distinguish two different targets which are not at least $10 \mu m$ far apart. A better accuracy should be achieved by repeating the measurement several times. Here, our beads sample is sufficiently sparse (beads are at least $>20 \mu m$ far apart) so that our localization technique distinguishes the three beads. Note that the more we go in depth, the more the fluorescence emission gets scattered, and the more the beads have to be far apart to still separate their respective fluorescent speckle envelopes. 

To show the consistency of the estimated positions based on centroid localization (in epi-detection, CAM1), we superimpose the real position of the beads (in bright field, CAM2) or refer to the positions of the focii (Fig.\ref{localizationv2}.B). Images are rescaled in microns by taking into account the magnification of our system (objective + tube lens) and the pixel size on both camera. 
In Fig.\ref{localizationv2}.C, we show that the estimated relative positions of the beads (based on fluorescent envelope localization, CAM1) are in good agreement with the ones we retrieve in with the control camera (CAM2). 
The time needed to focus on different beads is the main limitation of this technique.
 
\section{Discussion}
To summarize, we developed a new all-optical mechanism that focuses light inside scattering media. This technique relies on the use of a non-linear feedback signal, the spatial variance of the fluorescent speckle. It ensures the generation of a single diffraction-limited focus. An important advantage of our approach is that it works with linear signal, such as 1-photon fluorescence, and can potentially be interesting for Raman imaging. 
Unlike recent works, no memory effect is required which makes our technique even applicable with very thick scattering media as long as we detect fluorescence with sufficient signal-to-noise to perform the optimization.

As in any optimization process, the main drawback is the relatively long timescale to perform all the iterations. In our experiment, each iteration is slowed down by the collection of the fluorescent speckle (few tens of ms). Approaches encoding spatial information with a bucket detector such as single pixel camera techniques might increase the amount of collected fluorescence which reduces the exposure time and speeds up the whole optimization process.
Also, increasing the laser excitation power is also a possibility but bleaching of the fluorescent targets becomes an issue since the optimization process forms a focus on the target. Controlling the laser power throughout the optimization would limit this effect.

Additionally, the exact position of the focus cannot be determined in advance. However, focusing on different targets by performing multiple optimizations is possible by changing the algorithm parameters. We finally exploit the optimized fluorescent speckles to retrieve information about the relative position of the beads based on the centroid localization of their envelopes.

\newpage
\renewcommand{\refname}{References} 

\newpage

\section*{Supplementary Materials}
\subsection*{Variation of the spatial standard deviation with respect to $\Phi$}
Below we derive the theoretical variation of the variance with respect to the phase $\Phi$ of a given SLM segment/input mode.
We denote $I_{fluo}(x,y)$ the 2-dimensional fluorescent speckle intensity in the imaging plane. $I_{fluo}$ is the incoherent sum of all the speckles generated by each individual target: $I_{fluo}(x,y,\Phi) = \sum I_{exc}^{(k)}(\Phi) i_k(x,y)$, where:
\\
$\cdot$ $i_k(x,y)$ is the spatial shape of the speckle emitted by the $k^{th}$ target
\\
$\cdot$ $I_{exc}^{(k)}(\Phi)=a_k \sin(\Phi+\theta)+c_k$ corresponds to the intensity of the illumination speckle on target $k$. 
\\
Then the variance of the fluorescent speckle, $var(I_{fluo}(x,y,\Phi))$ reads:
\begin{equation}
\begin{split}
\small
\Var(I_{fluo}(x,y,\Phi)) 
& = \Var(\sum I_{exc}^{(k)}(\Phi) i_k(x,y)) \\
& = \sum I_{exc}^{(k)}(\Phi)^2 \Var(i_k(x,y)) \\ 
& = \sum (a_k\sin(\Phi+\theta)+c_k)^2 \Var(i_k(x,y)) \\
& = \sum (a_k^2\sin^2(\Phi+\theta)+2a_kc_k\sin(\Phi+\theta)+c_k^2)\Var(i_k(x,y)) \\
& \equiv A\sin(2\Phi+\theta_A)+B\sin(\Phi+\theta_B)+C
\end{split}
\label{eq2}
\end{equation}
where $A, \theta_A, B, \theta_B$ and $C$ are constants.

We explicitly see that the non-linearity introduced by the variance is of order 2. By comparison, the total 2-photon intensity, $I^{2p}(\Phi) = \sum \lambda_k^2(\Phi)$, presents the same evolution with respect to $\Phi$.
This explains why using the variance of our linear signal as a feedback enables us focusing light on a single target, like in 2-photon optimization.

\begin{figure}[htbp]
\centering
\includegraphics[width=300 pt] {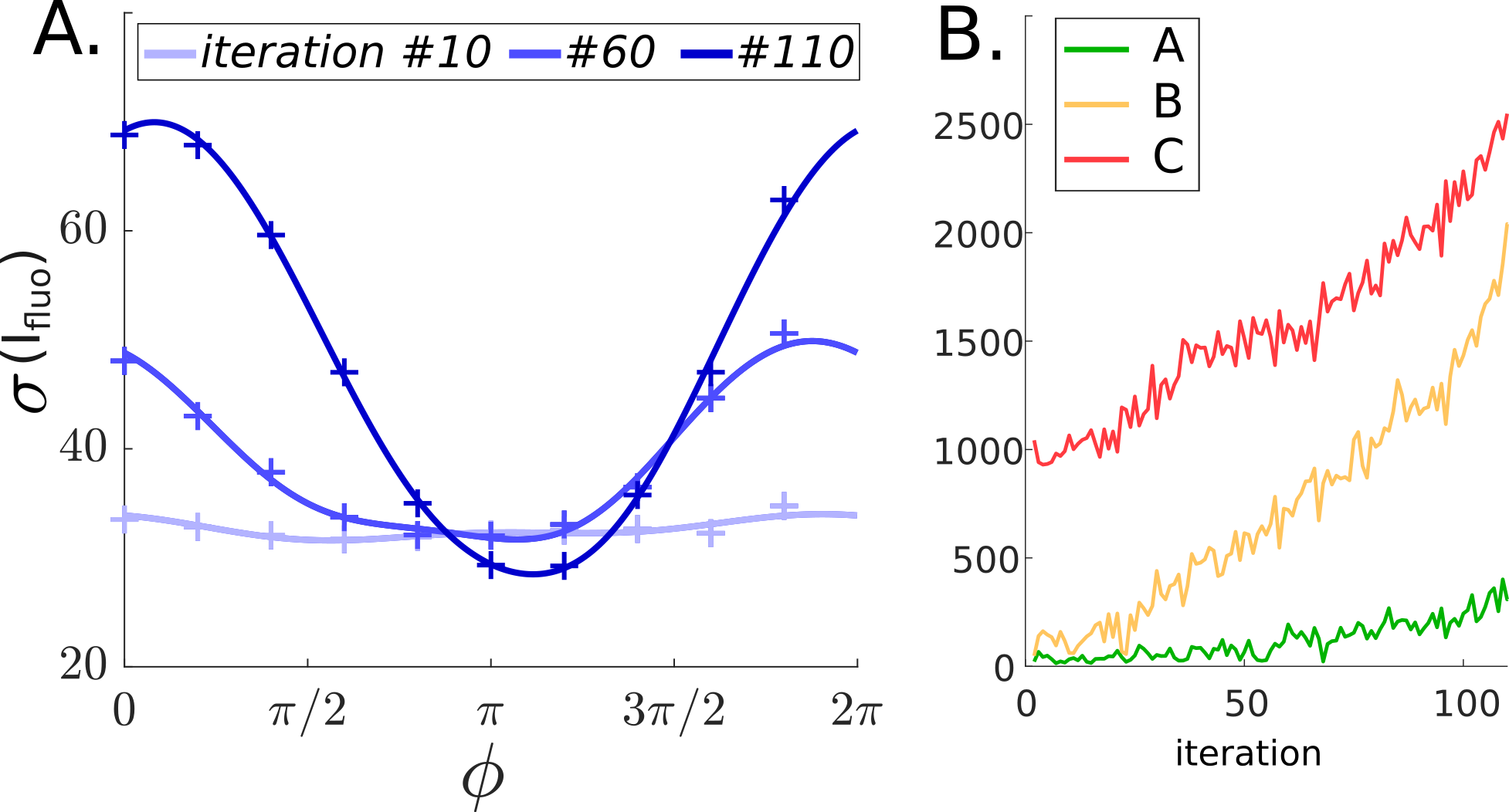}
\caption{\textbf{A}. Fit of standard deviation experimental data for three different Hadamard modes. \textbf{B}. Evolution of the corresponding fitting coefficients throughout the optimization.}
\label{modulation_phi}
\end{figure}

In Fig.\ref{modulation_phi}.A we validate our model and show that standard deviation experimental data, $ \sigma(I_{fluo})$, fit well with $\sqrt{eq.(\ref{eq2})}$. These curves also highlight the non-linearity of order 2 that is predominant for the first iterations (iteration $\#$10 and $\#$60). After a higher number of iteration, the fluorescence that we capture is predominantly emitted by a single target, which explains the sinusoidal shape.
Additionally, we look into the contribution of second and first order term in (\ref{eq2}), coefficient A and B respectively (Fig.\ref{modulation_phi}.B). The first iterations are the most non-linear ones with respect to $\Phi$: coefficients A and B are of the same order. Then, B is increasingly larger than A, which is consistent with the fact that the optimization mainly increases the total intensity rather than the contrast.

This theoretical model can also be used to estimate with higher accuracy the phase that maximize the variance.
This phase is estimated by performing $N_{step}$ measurement (corresponding to $N_{step}$ different phase masks displayed onto the SLM). Our model (eq. (\ref{eq2})) contains 5 independent parameters which imposes a minimal value for $N_{step}$ to correctly fit our experimental data. But after only few tens of iterations the coefficient A corresponding to the second order term is small compare to the first order one B. Therefore our fitting function can be approximated to $\sigma(I(x,y,\Phi)) \simeq \sqrt{(B\sin(\Phi+\theta_B)+C})$. Now we only need to determine 3 parameters, allowing us to decrease $N_{step}$, which speeds up the optimization procedure.

\subsection*{Performance for a larger number of beads}
\begin{figure}[htbp]
\centering
\includegraphics[width=300 pt] {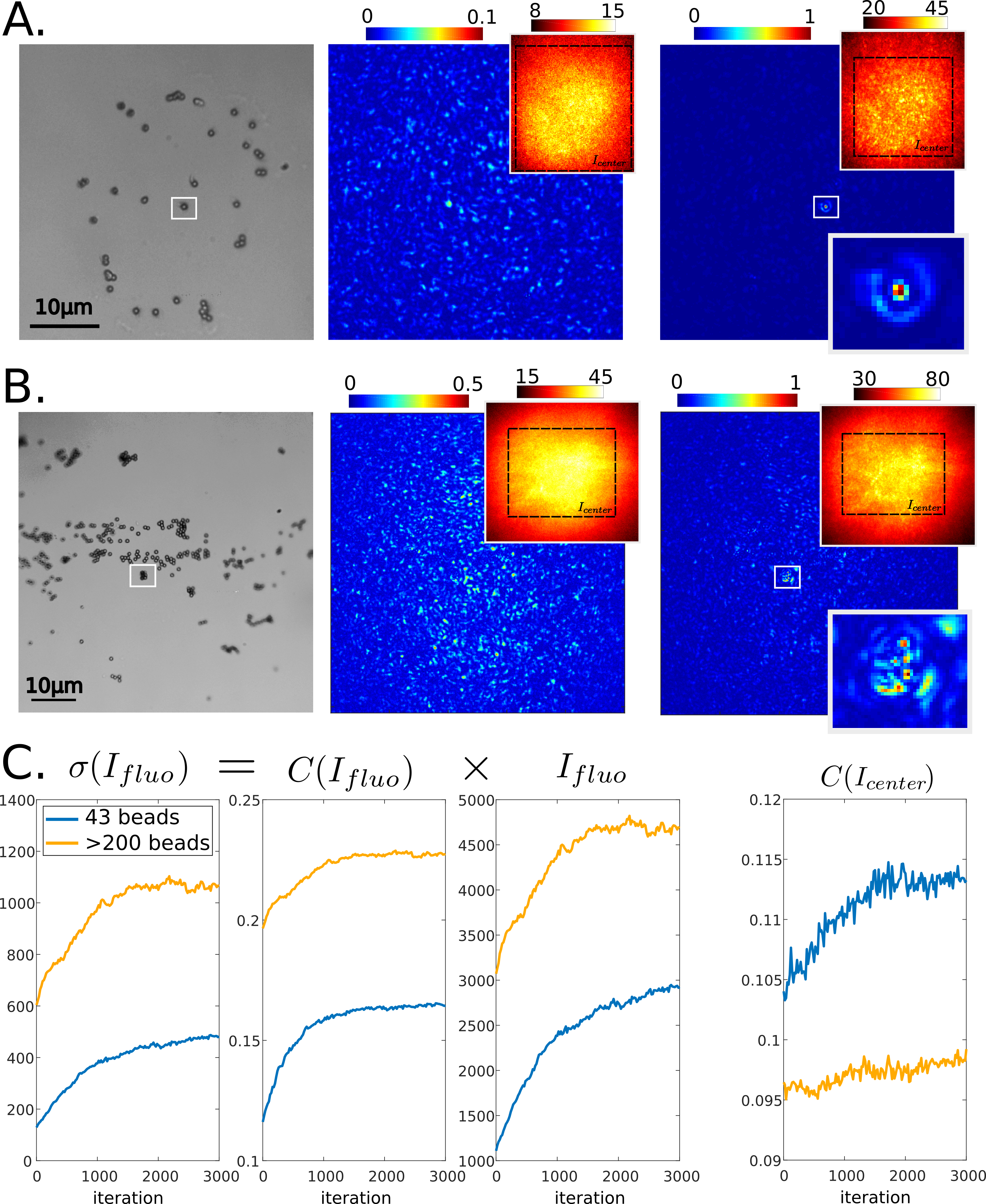}
\caption{Influence of the number of beads. \textbf{A}. With 43 beads a focus is still formed consistently at the end of the variance optimization. However when the number of targets become  too large (\textbf{B}) and the density increases, although the variance can still be optimized, as seen in (\textbf{C}), a focus is not always formed.}
\label{lotsofbeads}
\end{figure}

To show the strength and limitations of our technique, we increase the number of beads in our fluorescent sample. In particular, we show that a focus is still achievable up to a certain limit. In Fig.\ref{lotsofbeads}.A. we have 43 fluorescent beads, and as one can see the optimization procedure still consistently converge to a single diffraction-limited focus.
Nevertheless, if the sample contains a much larger number of beads (>200) with clusters (see Fig.\ref{lotsofbeads}.B) the illumination tends to form an extended focus. With such a number of fluorophores, the initial contrast is very low and the optimization seems to mainly enhance the contrast of the envelope ($C(I_{center})$ hardly increases, Fig.\ref{lotsofbeads}.C), thus focusing light on an extended area. However, this does not mean that our technique fails. An increase in standard deviation is noticeable. In this regime, the initial contrast can be enhanced by using a bandpass filter and/or an analyser at the cost of a loss of fluorescence. 

\end{document}